\documentclass[11pt,a4paper]{article}
\usepackage{amssymb} \usepackage{amsmath} \usepackage{graphicx}
\usepackage{epsfig,latexsym}
\usepackage{tabularx}

\def\bef{\begin{figure}}
\def\eef{\end{figure}}

\newcommand{\be}[1]{\begin{equation}\label{#1}}
\newcommand{\beq}{\begin{equation}}
\newcommand{\ee}{\end{equation}}
\newcommand{\beqn}[1]{\begin{eqnarray}\label{#1}}
\newcommand{\eeqn}{\end{eqnarray}}
\newcommand{\bd}{\begin{displaymath}}
\newcommand{\ed}{\end{displaymath}}

\def\lsim{\raise0.3ex\hbox{$\;<$\kern-0.75em\raise-1.1ex
e\hbox{$\sim\;$}}}  
\def\gsim{\raise0.3ex\hbox{$\;>$\kern-0.75em\raise-1.1ex
\hbox{$\sim\;$}}}
\def\simlt{\mathrel{\lower2.5pt\vbox{\lineskip=0pt\baselineskip=0pt
           \hbox{$<$}\hbox{$\sim$}}}}
\def\simgt{\mathrel{\lower2.5pt\vbox{\lineskip=0pt\baselineskip=0pt
           \hbox{$>$}\hbox{$\sim$}}}}
\def\unity{{\hbox{1\kern-.8mm l}}}

\newcommand{\vect}[1]{\mbox{\boldmath$#1$}}

\normalsize
\begin{document}

\baselineskip=0.65cm

\begin{center}
\Large
{\bf DAMA annual modulation effect and asymmetric mirror matter}
\rm
\end{center}

\large

\begin{center}

A. Addazi$^{a,b}$, Z. Berezhiani$^{a,b}$, R. Bernabei$^{c}$, P. Belli$^{c}$, F. Cappella$^{b}$, 
R. Cerulli$^{b}$, A. Incicchitti$^{d}$

\vspace{2mm}

$^{a}${\it Dipartimento di Scienze Fisiche e Chimiche,
 Universit\`a di L'Aquila, I-67100 Coppito AQ, Italy}\\
$^{b}${\it INFN, Laboratori Nazionali del Gran Sasso, I-67010 Assergi AQ, Italy}\\
$^{c}${\it Dipartimento di Fisica, Universit\`a di Roma ``Tor Vergata'' and \\ 
INFN -- Tor Vergata, I-00133 Rome, Italy}\\
$^{d}${\it Dipartimento di Fisica, Universit\`a di Roma ``La Sapienza'' and \\
INFN -- Roma, I-00185 Rome, Italy}\\

\vspace{1mm}

\end{center}
	
\normalsize

\begin{abstract}
The long-standing model-independent annual modulation effect measured by DAMA Collaboration 
is examined in the context of asymmetric mirror dark matter, assuming that dark atoms  interact with 
target nuclei  in the detector via kinetic mixing between mirror and  ordinary photons,  
both being massless.  
The relevant ranges for the kinetic mixing parameter are obtained taking into account various existing 
uncertainties in nuclear and particle physics quantities as well as
characteristic density and velocity distributions of dark matter in different halo models.

\end{abstract}

\vspace{5.0mm}


\vspace{2.0mm}


\section{Introduction}

Annual-modulation effect, as expected from the relative motion of the Earth 
with respect to the relic particles responsible for the Dark Matter (DM) in the galactic halo \cite{Drukier,Freese},
has been measured by DAMA Collaboration using the highly radiopure NaI(Tl) detectors of the former DAMA/NaI 
\cite{prop,allDM1,allDM2,allDM3,allDM4,allDM5,allDM6,allDM7,allDM8,Nim98,Sist,RNC,ijmd,ijma,epj06,ijma07,chan,wimpele,ldm,allRare1,allRare2,allRare3,allRare4,allRare5,allRare6,allRare7,allRare8,allRare9,allRare10,IDM96}
and of the second generation DAMA/LIBRA \cite{perflibra,modlibra,modlibra2,modlibra3,bot11,pmts,mu,review,papep,cnc-l,IPP,diu,norole} apparata.  
Measurements lasting for 14 annual cycles with an increasing exposure which cumulatively is equivalent to 
1.33 ton $\times$ year confirm  
the annual modulation effect at a confidence level of 9.3$\sigma$ \cite{modlibra3}.
No systematic effect or side reaction which could mimic the exploited signature, 
i.e. which would be able to account for the whole modulation amplitude and simultaneously 
satisfy all of the many peculiarities of the signature, has been found by 
the collaboration itself, neither it was suggested by anyone else over more than a decade.

The annual-modulation effect measured in DAMA experiments is model-independent. 
In other words, the annual modulation of the event rate is an experimentally established fact, 
independent on theoretical interpretations of the identity of dark matter and specifics of its interactions.
It can be related to a variety of interaction mechanisms of DM particles with the 
detector materials (see for example Ref. \cite{review}).  
In this paper we limit our analysis to the case where the signal is induced by atomic-type  dark matter 
candidates from asymmetric mirror sector.

Nowadays the concept that  DM may come from a hidden (or shadow)  
gauge sectors which have particle and interaction content similar to that of 
ordinary particles becomes increasingly popular.  
Such a dark sector may consist of elementary leptons  
(analogues of our electron) and  baryons  
(similar to our proton or neutron) composed of shadow quarks which are 
confined by strong gauge interactions like in our QCD. 
These two types of particles can be combined in atoms by electromagnetic forces 
mediated by dark photons. 
The stability of the dark proton is guaranteed by the conservation law of the related baryon number, 
as the stability of our proton is related to the conservation of the ordinary baryon number.  
On the other hand, the cosmological abundance of DM in the Universe 
can be induced by the violation of such baryon number in the early Universe 
which could produce dark baryon asymmetry by mechanisms similar 
to those considered for the primordial baryogenesis in the observable sector.  
In this respect, such type of DM is also known as asymmetric dark matter. 

Historically, the simplest model of such shadow sector, coined as mirror world, 
was introduced long time ago by reasons not much related to dark matter:  
it was assumed that there exists a hidden sector of particles 
that it is exactly identical to the ordinary particle sector, modulo parity transformation: 
for our particles being left-handed, the {\it parity} can be interpreted as a discrete 
{\it mirror} symmetry which exchanges them with their right-handed twins from parallel mirror sector 
\cite{Mirror1,Mirror2,Mirror4,Mirror5}.  
Thus, all ordinary particles described by the Standard Model 
$SU(3)\times SU(2)\times U(1)$,  
the electron $e$, proton $p$, neutron $n$, photon $\gamma$, neutrinos $\nu$ etc., 
should have mass-degenerate invisible twins: $e'$, $p'$, $n'$, $\gamma'$, $\nu'$ etc.\ 
which are sterile to our strong, weak and and electromagnetic interactions  
but have instead their own gauge interactions,  
described by the mirror copy of the Standard Model 
$SU(3)'\times SU(2)'\times U(1)'$ with exactly the same quantum numbers and coupling constants.  
In this case, the microphysics of the mirror matter should be exactly the same as that of ordinary matter, 
at all levels from particle to atomic physics. 

Naively, then one would  expect that mirror world would be 
identical to the observable world also in cosmology, which situation could be disfavored 
by the following simple arguments: 
 
(a)  once mirror baryon asymmetry is generated 
by the same physics as the ordinary one, then the cosmological density of the mirror 
baryons in the Universe was expected to be the same as that of the ordinary baryons, 
$\Omega'_B = \Omega_B$. This would not be sufficient for explaining the whole amount of DM.  

(b) the mirror sector with the same abundances of mirror photons and neutrinos as the ordinary one  
would strongly disagree with the Big Bang Nucleosynthesis (BBN) limits 
on the effective amount of light degrees of freedom. It would be equivalent to effective 
amount of extra neutrinos $\Delta N_{\rm eff} = 6.15$. 

By these reasons, and also because of self-interacting and dissipative nature of the mirror 
matter,  it was not considered  as a serious candidate for dark matter for a long time. 
However, as it was shown in \cite{Berezhiani:2000gw}, all problems can be avoided 
assuming that after inflation the two sectors were heated to different temperatures, 
and the temperature of the mirror sector $T'$ remained less than the ordinary one $T$ 
over all stages of the cosmological evolution. 
The BBN limits are satisfied if $T' < T/2$ or so,  while for  $T' < T/4$ mirror matter 
can represent the entire fraction of DM since in this case   
early decoupling of the mirror photons renders mirror baryons practically 
indistinguishable from the canonic cold dark matter (CDM) in observational tests 
related to the large scale structure formation and CMB anisotropies  
\cite{Berezhiani:2000gw,Ignatiev,Berezhiani:2003wj}. 
Interestingly, the condition $T' < T$ can also lead to mirror baryon asymmetry  
bigger than the ordinary one,  
and $\Omega'_B \simeq 5 \Omega_B$ can be naturally obtained   
in certain co-baryogenesis scenarios  
\cite{Bento:2001rc,Bento:2002sj,Berezhiani:2005hv,Berezhiani:2003xm}.  
Therefore, the mirror matter can be a viable candidate for DM, 
despite its collisional and dissipative nature, and the 
present situation in fundamental physics and cosmology gives new perspectives for testing 
this intriguing hypothesis with a rich predictive power 
(for reviews, see e.g. \cite{Berezhiani:2003xm,Foot:2003eq,Berezhiani:2005ek}).  

More in general, the mirror DM  can also be presented in the form known as asymmetric mirror matter,  
which assumes that mirror parity is spontaneously broken and  
the electroweak symmetry breaking scale $v'$ in the mirror sector is much larger than that 
in our Standard Model,  $v = 174$ GeV
\cite{Berezhiani:1995yi,Berezhiani:1995am,Berezhiani:1996sz,Mohapatra:1996yy}. 
In this case, the mirror world becomes a heavier and deformed copy of our world, 
with mirror particle masses scaled in different ways with respect to the masses of the ordinary particles.   
Taking the mirror weak scale e.g. of the order of 10 TeV,   
the mirror electron would become two orders of magnitude heavier than our electron       
while the mirror nucleons $p'$ and $n'$  only about 5 times heavier than the ordinary nucleons. 
Then dark matter would exist in the form of mirror hydrogen 
composed of mirror proton and electron, with mass of about 5 GeV 
which is a rather interesting mass range for dark matter particles  
\cite{Berezhiani:2006ac,Berezhiani:2008zza}. 
In the context of baryo-cogenesis mechanisms 
\cite{Bento:2001rc,Bento:2002sj,Berezhiani:2005hv,Berezhiani:2003xm}, 
$\Omega'_B \simeq 5 \Omega_B$ can be simply related to  
the mass ratio $\simeq 5$ between mirror and ordinary baryons  
\cite{Berezhiani:2003xm,Berezhiani:2006ac,Berezhiani:2008zza} 
(see also Ref. \cite{An:2009vq}). 
Owing to the large mass of mirror electron, mirror atoms should be more compact 
and tightly bound with respect to ordinary atoms. 
Asymmetric mirror model can be considered as a natural benchmark for more generic
types of atomic dark matter with {\it ad hoc} chosen parameters. 

In this paper we discuss the annual modulation observed by DAMA in the framework of 
asymmetric mirror matter, in the light of the very interesting interaction portal which is 
kinetic mixing $\frac{\epsilon}{2} F^{\mu\nu} F'_{\mu\nu}$  of two massless states, 
ordinary photon and mirror photon \cite{Holdom,Mirror5}. 
This mixing mediates the mirror atom scattering off the ordinary target nuclei in 
the NaI(Tl) detectors at DAMA/LIBRA set-up with the Rutherford-like cross sections. 
The compactness of mirror atoms is important here. 
On the one hand, if these atoms are sufficiently fuzzy,  
the electromagnetic scattering between ordinary and mirror nuclei 
will not be suppressed by the mirror atom form-factor. 
On the other hand, they should be rather compact to render 
self-interaction of the mirror matter weak enough -- otherwise the structure of galactic halo would be 
strongly affected. Moreover, if the scattering cross-section between mirror atoms is such that 
$\sigma/M \sim 10^{-24}-10^{-23}~{\rm cm}^2/{\rm GeV}$, where $M$ is the dark atom mass, 
some characteristic problems for canonic CDM models as the cusp problem 
or overly large number of small halos within the local group can be avoided  
\cite{Spergel:1999mh,Wandelt:2000ad}. 

The paper is organized as follows. In Sect. 2 we give a brief overview of asymmetric 
mirror dark matter models discussing its direct detection via photon-mirror photon kinetic mixing. 
In Sect. 3 details of the analysis are given,
while in Sect. 4 we discuss the obtained results.

\section{Asymmetric mirror matter and its direct detection}

\subsection{Overview of asymmetric mirror matter} 

At the basic level, ordinary sector is described by the Standard Model 
$G_{\rm SM} = SU(3) \times SU(2) \times U(1)$ containing 
quarks $q=(u,d)$ and leptons $l=(\nu,e)$  of three generations  and 
the Higgs doublet $H$.  Then the dark mirror sector must be described by the 
identical gauge group $G'_{\rm SM} = SU(3)' \times SU(2)' \times U(1)'$  
containing mirror quarks and leptons, $q'=(u',d')$ and $l'=(\nu',e')$,   
and mirror Higgs $H'$.
The most general Lagrangian of such $G_{\rm SM}  \times G'_{\rm SM}$ theory has the form 
\begin{equation}\label{L}
{\cal L}_{\rm tot} = {\cal L} + {\cal L}' + {\cal L}_{\rm mix}
\end{equation} 
where the Lagrangians ${\cal L}= {\cal L}_{\rm gauge} + {\cal L}_{\rm Yuk} + {\cal L}_{\rm Higgs}$ 
and ${\cal L}'= {\cal L}'_{\rm gauge} + {\cal L}'_{\rm Yuk} + {\cal L}'_{\rm Higgs}$ 
describing all interactions in the ordinary and mirror sectors can be rendered identical 
by imposing mirror parity, a discrete symmetry that exchanges ${\cal L} \leftrightarrow {\cal L}'$.    
The ${\cal L}_{\rm mix}$ describes the possible interactions between ordinary and mirror 
particles as e.g. photon-mirror photon kinetic mixing  which shall be discussed later. 
One can make a further step extending towards the concepts of supersymmetry 
(SUSY) and grand unification theory (GUT).
In view of the gauge coupling unification, 
envisaging that both sectors, ordinary and dark, in the ultraviolet limit are described 
by a SUSY GUT $G\times G'$ with two identical gauge factors that   
can be $SU(5)$, $SO(10)$  or $SU(6)$. 
Then both gauge factors $G$ and $G'$  are spontaneously broken down to their 
standard subgroups at the grand unification scale around $10^{16}$ GeV. Below this scale 
ordinary and mirror sectors are represented by respective Standard Models 
(or SUSY Standard Models),     
$SU(3) \times SU(2) \times U(1)$ and $SU(3)' \times SU(2)' \times U(1)'$, 
with identical particle contents and identical patterns of interaction constants 
(gauge and Yukawa).\footnote{In the following, we do not consider supersymmetric 
contributions to dark matter. The SUSY DM in the form of neutralinos  
can be destabilized by introducing a tiny violation of $R$-parity, or perhaps not so tiny but 
in specific forms which would not affect the matter stability itself, as e.g. in Refs. 
\cite{Barbieri:1997zn,Giudice:1997wb}.}  
In the case when the mirror parity remains unbroken at all levels, 
so called exact mirror model, we have no new parameter: 
ordinary and mirror particles are degenerate in mass, 
and the microphysics of mirror and ordinary sectors is exactly the same.   

However, one can consider another possibility
when mirror parity is spontaneously broken at the level of electroweak interactions 
so that the vacuum expectation value (VEV) of mirror Higgs $\langle H' \rangle = v'$ is much larger than 
that of the ordinary Higgs $\langle H \rangle = v \approx 174$~GeV   
\cite{Berezhiani:1995yi,Berezhiani:1995am,Berezhiani:1996sz}.\footnote{Mechanisms 
of discrete symmetry breaking between ordinary and mirror sectors were discussed 
e.g. in Refs.  \cite{Berezhiani:1995am,Berezhiani:1996sz}. 
Seemingly, there emerges a hierarchy problem between the two VEVs $v'$ and $v$, 
which however can be turned into positive aspect ameliorating the hierarchy problem for 
the ordinary Higgs, in the context of supersymmetric little 
Higgs or twin Higgs models \cite{Berezhiani:2005ek,Barbieri:2005ri,Chacko:2005vw}. 
In these models the Higgs in ordinary sector emerges as a pseudo-Goldstone boson 
associated with the breaking of the accidental global $SU(4)$ symmetry between two 
Higgs systems, and thus the mass and VEV of ordinary Higgs $H$ can be smaller, 
within two orders of magnitude or so, than the VEV of the mirror Higgs $H'$, $v' \sim 10$~TeV. 
}
Such a concept of asymmetric mirror matter  introduces a new parameter,  
the ratio $\zeta= v'/ v$  between the VEVs, 
while the gauge and Yukawa constants in two sectors essentially remain identical. 
 When $\zeta \gg 1$, the mirror sector becomes a deformed heavier copy of the ordinary sector, 
with mirror particle masses scaled in different ways as functions of $\zeta$
with respect to the masses of their ordinary partners.    
Namely, the masses of  shadow quarks and charged leptons are 
induced by the Yukawa couplings with the Higgs $H'$, and so they scale roughly 
by a factor $\zeta$ with respect of the masses of their ordinary twins
(e.g. $m'_{e,u,d} \simeq \zeta  m_{e,u,d}$ for the electron, 
up-quark and down-quark, if one neglects the renormalization group running factors 
for quarks from the scale $\langle H'\rangle$ down to $\langle H\rangle$), 
the mirror neutrino masses which emerge from the dimension-5 operators 
$\frac{1}{M} l'l' H'H'$ quadratic in Higgs field, scale by a factor $\zeta^2$ with respect 
to ordinary neutrino masses generated by similar operators $\frac{1}{M} ll HH$. 
In addition, the `Fermi' constant of mirror weak interactions becomes smaller by a factor $\zeta^2$, 
with respect to our Fermi constant $G_F$, while the infrared scale of mirror QCD 
changes roughly as $\Lambda' \sim \zeta^{0.28} \Lambda$ with respect to our QCD scale
due to the threshold effects of heavier mirror quarks.  
If $\zeta \leq 10^2$, for the masses of light mirror quarks $u'$ and $d'$ we have  
$m'_{u,d} < \Lambda'$, and so for the mass of mirror pions we get roughly $m'_\pi/m_\pi \sim \zeta^{0.6}$. 
Hence, taking e.g. $\zeta =100$, 
shadow electrons become two orders of magnitude heavier than our electron,      
shadow neutrinos $10^4$ times heavier than our neutrinos, 
but shadow proton and neutron only about 5 times heavier than ordinary nucleons, 
e.g. $m'_p \simeq \zeta^{0.28} m_p + \zeta (2 m_u + m_d)$,   
due to larger  by larger QCD scale and larger masses of light mirror quarks.  
Hence, mirror nucleon masses become  about 5 GeV 
which is an interesting mass range for dark matter particles. 
In addition, due to the large mass gap between the masses of mirror neutron and proton, 
$m'_n - m'_p \gg m'_e$, the neutron becomes unstable even if it is bounded in nuclei, 
and hence the only stable atom in such shadow sector can be the mirror hydrogen 
composed of mirror proton and electron 
\cite{Berezhiani:1995yi,Berezhiani:1995am,Berezhiani:1996sz}. 
Needless to say, that mirror hydrogen can be presented in molecular form, while some 
its fraction can exist in ionized state: if dark electron and proton are sufficiently 
heavy, the matter may remain essentially ionized owing to  very small recombination 
cross section. 

The baryogenesis in the two sectors, ordinary and mirror, emerges by the same mechanism, 
since the particle physics responsible for baryogenesis is the same in the two sectors 
(coupling constants, CP-violating phases, etc.).  
The spectrum of mirror particle masses is irrelevant for the baryon asymmetry
in the mirror sector as far as the effective scale of interactions relevant for baryogenesis 
is much higher than the mirror electroweak scale. 
However, the cosmological conditions at the baryogenesis epoch can be 
different (recall that shadow sector must be colder than ordinary one). 
One can consider two cases: 

1) {\it Separate baryogenesis}, when the baryon asymmetry in each sector is generated 
independently but by the same mechanism. 
In this case, in the most naive picture when out-of-equilibrium conditions are well satisfied 
in both sectors, one predicts $\eta = n_B/n_\gamma$ and $\eta' = n'_B/n'_\gamma$ 
must be equal, while $n'_\gamma/n_\gamma \simeq x^3 \ll 1$, where $x = T'/T$ is 
the temperature ratio between mirror and ordinary worlds in the early Universe.   
In this case, we have $\Omega'_B/\Omega_B \simeq (m'_N/m_N) x^3 $ where 
$m'_N$ is the mass of shadow nucleon vs. the mass of ordinary nucleon $m_N\simeq 1$~GeV.  
Therefore, if e.g. $x = 0.5$,  for 
$\Omega'_B/\Omega_B \simeq 5$  we need $m'_N = 40$ GeV. 
In the context of asymmetric rescaling of particle masses, this would occur 
when $\zeta \sim 2000$ or so, in which case for mirror electron mass one would have 
$m'_e \sim 1$ GeV or so. In this case, we are obviously in rather heavy range of dark matter. 
However, one should remark that  
due to different out-of equilibrium conditions in two sectors 
situation with $\eta' \gg \eta$ can be also obtained in some specific parameter space 
\cite{Berezhiani:2000gw}.

2) {\it Co-genesis} of baryon and mirror baryon asymmetries via $B-L$ and 
CP-violating processes between the ordinary and mirror particles, e.g. 
by the terms $ \frac{1}{M}\, l l' H  H' $ in ${\cal L}_{\rm mix}$ which also 
 induce mixing between ordinary (active) and mirror (sterile) neutrinos,  
 and which can be mediated by heavy ``right-handed'' neutrinos  coupled to 
both sectors as e.g. \cite{Bento:2001rc,Bento:2002sj}.  
This leptogenesis mechanism predicts $n'_B = n_B$ and so for  
$\Omega'_B/\Omega_B \simeq 5$ we need $m'_N/m_N \simeq 5$, which 
singles out the mass of dark atom of about 5 GeV. (Somewhat different leptogenesis 
mechanism also based on the assumption $m'_N > m_N$  
was suggested in Ref. \cite{An:2009vq}.)
This would occur when $\zeta \sim 100$ 
in which case the mirror electron mass is $m'_e \sim 50$ MeV 
\cite{Berezhiani:2003xm,Berezhiani:2005ek,Berezhiani:2006ac,Berezhiani:2008zza}.

Let us remark, however, that in the picture of asymmetric mirror world 
the mirror (sterile) neutrinos could also be a natural candidate of dark matter, 
namely warm dark matter. 
In fact, assuming a minimal mass normal hierarchy for ordinary neutrino masses, 
one can approximate to current cosmological data with a single mass eigenstate with 
mass $m_\nu = 0.06$~eV. We have then 
$\Omega_\nu h^2 \approx \sum m_\nu/93~{\rm eV} \approx 6 \times 10^{-4}$, 
well below the cosmological limits on the neutrino masses.  
Since the mirror neutrino masses scale roughly as $m'_{\nu_a} = \zeta^2 m_{\nu_a}$, 
$a=1,2,3$, 
then for $\zeta \sim 10^2$ or  more we would have 
$\sum m'_\nu \approx  \zeta^2 \sum m_\nu \approx (\zeta/100)^2 \times 0.6$~keV 
and $\Omega'_\nu h^2 \approx \zeta^2 x_\nu^3 \sum m_\nu/93~{\rm eV} 
\approx 0.1 (\zeta/100)^2 (x/0.25)^3 $, where $x_\nu = T'_\nu/T_\nu$ is the temperature 
ratio between the mirror and ordinary neutrinos in the early Universe (which can be 
somewhat smaller than the photon temperature ratio $x=T'/T$ if $\zeta$ is large)
\cite{Berezhiani:1995am,Berezhiani:1996sz}.
Therefore, for $\zeta \simeq 100$ and 
$x\simeq 1/4$ practically the whole budget of dark matter will be saturated by mirror neutrinos 
with mass $\sim 1$~keV leaving practically no space for mirror baryons. 
Therefore, in the following we assume that the mirror sector is cold enough to leave 
a space for mirror baryons as well.  
In any case, in what follows, we do not require that mirror baryons 
provide entire amount of DM, but  we assume that it provides some fraction $f$ 
of DM which we shall keep as an arbitrary parameter. 

How large this fraction can be depends on the mass spectrum of the mirror particle 
and it is limited by the degree of self-interaction of mirror atoms.  The self-scattering cross 
section of dark matter particles should satisfy the (most conservative) upper limit
$\sigma/M < 10^{-23}~{\rm cm}^2/{\rm GeV}$ \cite{Spergel:1999mh,Wandelt:2000ad}.  
For this, mirror atoms should be enough compact,  
with small enough mirror Bohr radius $a' = (\alpha' m'_e)^{-1} \simeq a/\zeta$
where we take $m'_e \ll m'_p$ and  
$a = 5.3 \times 10^{-9}$~cm is ordinary Bohr radius. 
In the case of mirror hydrogen this condition cannot be easily satisfied. The hydrogen atom has 
a long tail potential which in fact is responsible for molecular hydrogen as a bound state 
of two atoms. Therefore, the scattering cross section is resonantly amplified when the 
energy of scattering $E$ gets close to the energy levels of the hydrogen molecule. 
One can parametrize this cross section as 
$\sigma(E) = F(E) a^{\prime 2}$. 
At energies exceeding the atom binding energy, $E \geq E'_0= \alpha^{\prime 2} m'_e/2$, 
one has $F(E) \leq 1$ , but for $E \ll E_0$ the cross-section strongly increases,  
and we have $F(E) \sim 100$ or so \cite{Cline:2013pca}. 
Let us consider e.g. the case $m'_p=6$~GeV for the dark proton mass 
which can be obtained when $\zeta \simeq 200$.  
Then at typical virial velocities $v \sim 300$~km/s characteristic for large galactic halos, 
we have $E \simeq 3$ keV, which is comparable to $E'_0$ 
when $m'_e = \zeta m_e \simeq 100$~MeV. 
However, in this case we get 
$\sigma/M \simeq a^{\prime 2}/m'_p \sim 10^{-22}~{\rm cm}^2/{\rm GeV}$
which exceeds by an order of magnitude the above upper limit 
($a^{2}/m_p \sim 3 \times 10^{-17}~{\rm cm}^2/{\rm GeV}$ for ordinary hydrogen). 
If we take instead $\zeta = 2000$, we get $m'_e \simeq 1$ GeV 
and $m'_p \simeq 20$~GeV or so. In this case we have $E \simeq 10$ keV against
$E'_0 \simeq 30$~keV,  so $F(E) \sim 10$ and 
$\sigma/M \simeq 10 a^{\prime 2}/m'_p \sim 3 \times 10^{-24}~{\rm cm}^2/{\rm GeV}$ 
which can be acceptable for large halos, with $v\simeq 300$~km/s. 
Moreover, such a self interaction would smooth out the cusp providing a central density 
of halos of about $10^{-2}~M_\odot/{\rm pc}^3 \simeq 0.3~{\rm GeV}/{\rm cm}^3$ 
\cite{Spergel:1999mh,Wandelt:2000ad}. 
However, for small halos, with e.g. $v\sim 30$~km/s,  
we would have $E\ll E_0$ and so $F(E) \geq 100$ \cite{Cline:2013pca}, 
and thus we would get 
$\sigma/M \geq 3 \times 10^{-23}~{\rm cm}^2/{\rm GeV}$, again larger than the above conservative 
upper limit. Therefore, in the cases discussed above mirror hydrogen cannot provide 
the entire amount of dark matter. In the case $\zeta \sim 10^2$ it can be only a subdominant 
fraction (e.g. $f \sim 0.1$) while the rest of dark matter can be represented e.g. by sterile mirror 
neutrinos with masses of few keV. 
In the case  $\zeta \sim 10^3$ mirror hydrogen could be a dominant fraction, 
say $f \simeq 0.8$ or so. However, it would be not effective for small dark matter halos 
(dwarf galaxies) and  thus the latter must be formed essentially by mirror neutrinos 
constituting a small fraction, say 20 per cent or so of the whole amount of DM. 
This transition of DM content from large to small halos 
may provide an interesting explanation to the reduced  
amount of small substructures and ``Too big to fail'' problems.   

More generically,  the low energy theory in both sectors could 
be represented by the Standard Model containing 
two Higgs doublets $H_u$ and $H_d$, as e.g. in the context of supersymmetry 
or in the model  providing an axion solution to the strong CP-problem \cite{Berezhiani:2000gh}.  
In this case the ratio of two VEVs $\tan\beta' = \langle H'_u \rangle/\langle H'_d \rangle$ 
in mirror sector could be different from $\tan\beta = \langle H_u \rangle/\langle H_d \rangle$. 
In other words, up and down quark masses in the mirror sector will scale 
by different factors $\zeta_u = \langle H'_u\rangle/\langle H_u \rangle$
and $\zeta_d = \langle H'_d\rangle/\langle H_d \rangle$: we have 
$m'_{e,\mu,\tau} = \zeta_d m_{e,\mu,\tau}$ for charged leptons and  
$m'_{d,s,b} = \zeta_d  m_{d,s,b}$ for down-type quarks, but 
$m'_{u,c,t} = \zeta_u m_{u,c,t}$ for up-type quarks and 
$m'_{\nu_e,\nu_\mu,\nu_\tau} = \zeta_u^2 m_{\nu_e,\nu_\mu,\nu_\tau}$ 
for the neutrino Majorana masses. 

For the time being, $\tan\beta$ is a free parameter for the MSSM or for the two Higgs model 
which formally can range from 1 to about 100 in ordinary sector. 
The case when $\tan\beta' = \tan\beta$, 
i.e. $\zeta_u = \zeta_d = \zeta$, was discussed above. 
However, if $\tan\beta' \neq \tan\beta$, other interesting options can appear. 
For example, taking $\zeta_u \sim 10^2$ and $\zeta_d \sim 10^3$, which correspond 
to the case e.g. $\tan\beta \simeq 10$ and $\tan\beta' \simeq 1$, 
we would have $m'_e \simeq \zeta_d m_e \simeq 500$~MeV. 
But now mirror upper quarks are rather light, 
$m'_u \sim 100 m_u \sim 250$ MeV, while mirror down quarks become 
very heavy $m'_d \sim 10^3 m_d \sim 5$~GeV. 
In this case down quarks are unstable against $\beta$-decay 
even when they are bounded in hadrons, and thus the lightest baryon in mirror sector 
should exist in the form of $\Delta^{\prime ++} = u'u'u'$ bound state with spin 3/2 and mirror 
electric charge +2, rather than in the form of the mirror proton. 
The mass of such baryon will be rescaled roughly as $\Lambda'/\Lambda \simeq 4$  
with respect to ordinary $\Delta$-resonances, $M_\Delta \simeq 1.2$~GeV. 
On the other hand, because of charge +2, it must form a helium-like atom with two electrons, 
having the mass $M'_A \sim 6$ GeV or so.  
In this case the chemical neutrality of the helium gives an interesting indication 
for the self-scattering cross section. Taking into account that the helium atom radius 
is about 3/5 of the Bohr radius $a'=a/\zeta_d$,  we obtain 
$\sigma_{A'A'}/M'_A \simeq 2 \times 10^{-24}~{\rm cm}^2/{\rm GeV}$ in 
excellent agreement with the self-scattering limits. Therefore, such exotic helium atom 
could constitute the entire dark matter in the galaxies and in the whole universe.  

There can emerge other interesting possibilities.  
E.g. for  a proper choice of $\tan\beta' \simeq 2 \tan\beta$,
mirror proton and mirror neutron can be enough degenerate in mass, 
both could be stable and also form mirror nuclei 
also with rather large atomic numbers. 
Alternatively, if $\tan\beta' > 2 \tan\beta$ or so, mirror proton 
could become heavier than mirror neutron and it can decay into the latter. 
In this case mirror  world will be represented solely by mirror neutrons with mass 
say 5 GeV and with strong self-interaction cross-section, 
$\sigma_{n'n'} \sim 10^{-24}$ cm$^2$ or so. Hence, we have 
rate $\sigma/M\sim 10^{-24}$ cm$^2$/GeV, which is in perfect agreement with all limits 
on self-scattering  DM and yet can solve the problems of galactic halo shape.   

\subsection{Photon-mirror photon kinetic mixing portal} 

The mirror matter can interact with ordinary matter through the {\it photon-mirror photon kinetic mixing}.
In the context of $G_{\rm SM}  \times G'_{\rm SM}$,
the kinetic mixing between gauge bosons of two abelian factors, 
$\frac{\tilde \epsilon}{2} \, B^{\mu\nu} B'_{\mu\nu}$ is allowed. 
After the electroweak symmetry breaking, it transforms into 
photon -- mirror photon kinetic mixing term 
\begin{equation}\label{eps}
 \frac{\epsilon}{2} \, F^{\mu\nu} F'_{\mu\nu} 
\end{equation} 
with  $\epsilon =  \tilde\epsilon \cos\theta_W \cos\theta'_W$. 
This mixing does not induce oscillation between ordinary and mirror photons 
as far as both are massless. However, it
in fact makes mirror particles mini-charged with respect to ordinary 
electromagnetic interactions: the mirror particles acquire electric charges $\epsilon q$. 
Generically, in this case dimensionless parameter $\epsilon$ could be order 1. 

On the other hand, there are stringent experimental constraints on the parameter 
$\epsilon$ which depend on the masses of mini-charged particles 
\cite{Mohapatra:1991as,Davidson:1993sj,Prinz:1998ua,Davidson:2000hf,Badertscher:2006fm,Berezhiani:2008gi}.  
As we see below, our results for dark matter detection are compatible with the existing 
limits on the dark matter particle mini-charges. 
In particular, regarding the mass range of mirror particles we are interested in, 
taking mirror electron mass e.g. $m'_e \simeq 100$ MeV, we have a direct  experimental   limit 
$\epsilon < 6 \times 10^{-4}$ \cite{Prinz:1998ua}, 
while cosmological BBN limit from $e^+e^-$ s-channel annihilation into mirror electron positron pair 
gives $\epsilon < 10^{-7}(m'_e/1~{\rm GeV})^{1/2}$ or so \cite{Berezhiani:2008gi}.  

There emerges a question (or naturalness problem), i.e. why $\epsilon$ is so small, 
which can be easily resolved by considering a grand unified version of mirror matter, 
$G \times G'$ with e.g. $G = SU(5)$ or $SU(6)$. 
In this case, 
no kinetic mixing is possible between the non-abelian gauge bosons of two sectors 
without the GUT symmetry breaking. 
Hence, the term (\ref{eps}) can emerge only from the higher dimensional operator 
$\frac{\kappa}{2M_P^2} \, (\Sigma G^{\mu\nu}) (\Sigma' G'_{\mu\nu})$ where 
$G^{\mu\nu}$ are gauge fields of the GUT, $\Sigma, \Sigma'$ are Higgs fields in adjoint 
representations which break the GUT symmetry down to the Standard Model, 
and $M_P$ is some natural cutoff scale, of order of Planck mass or so. 
Therefore, photo-mirror photon kinetic mixing can emerge 
only after breaking of the GUT symmetries by the VEVs 
$\langle \Sigma \rangle = \langle \Sigma' \rangle \sim 10^{15-16}$ GeV, and so one can 
naturally have  $\epsilon < 10^{-7}$ even if constant $\kappa$ is order of one
\cite{Berezhiani:2003xm,Berezhiani:2005ek}.

\begin{figure}[t]
\centerline{\includegraphics [width=5cm]{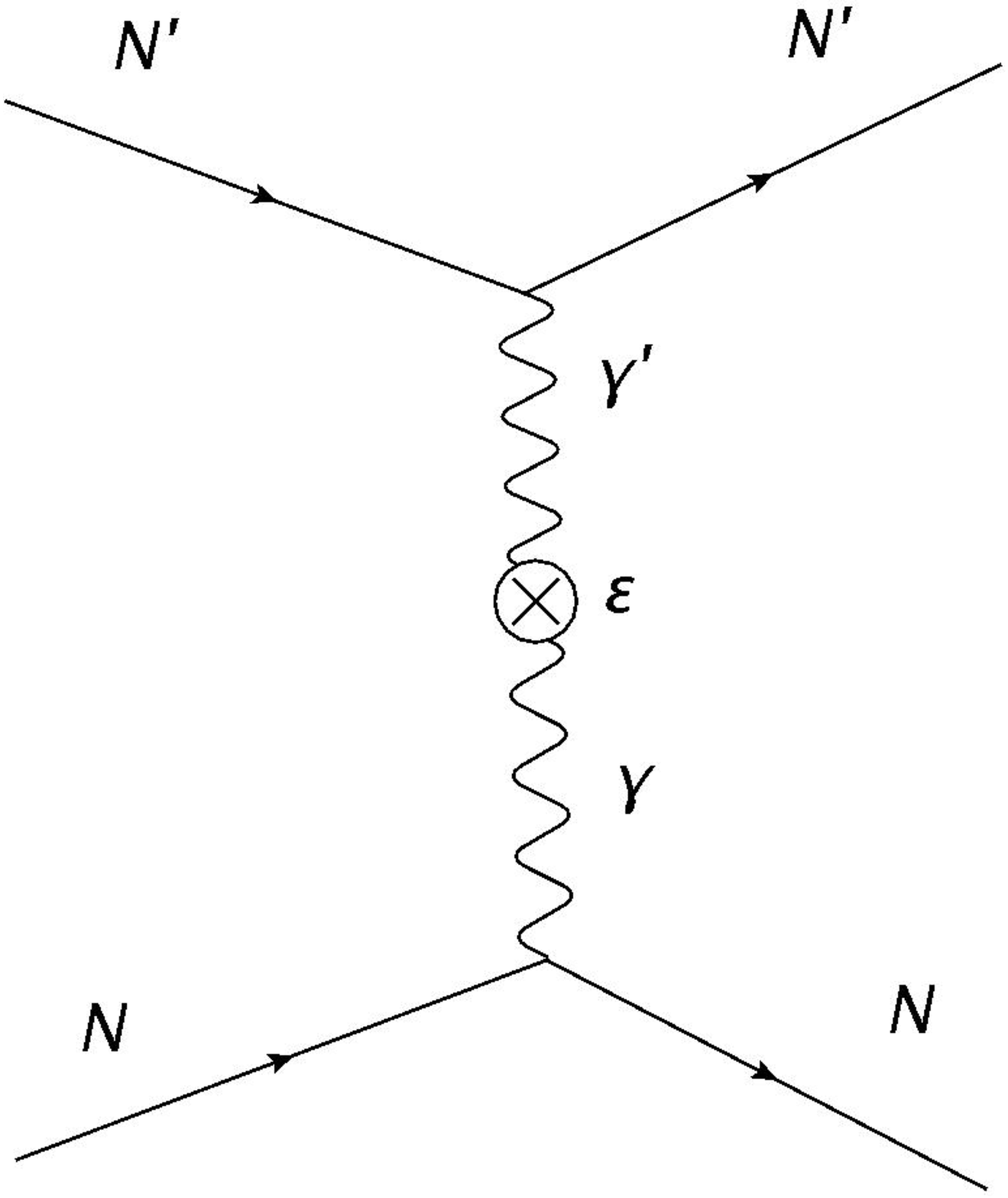} }
\caption{Mirror Nucleus-Nucleus interaction through Photon-Mirror photon portal. }
\label{plot1}   
\end{figure}

The basic mechanism in the mirror atom scattering off the ordinary target 
is the following\footnote{In the case of exact mirror model, 
this mechanism was discussed by Foot \cite{Foot:2003iv,Foot:2012ty}. }:  
the mirror nuclei $\mathcal{N}'$, with 
a mirror electric charge $Z'$, interacts with $\mathcal{N}$ ordinary nuclei with
the electric charge $Z$:
\begin{equation}\label{NN}
\mathcal{N}'+\mathcal{N} \rightarrow \mathcal{N}'+\mathcal{N}
\end{equation}
via kinetic mixing of two photons as shown in Fig. \ref{plot1} 
with Rutherford-like cross section suppressed by the free parameter $\epsilon \ll 1$. 
(The direct scattering of the mirror nuclei scattering off the electrons will be irrelevant by
obvious reasons). 
The mirror nuclei can be simply a mirror proton, $Z'=1$, in the case when 
mirror world is dominated by hydrogen-like atoms. In other situation discussed above, 
when mirror matter is presented by $\Delta$-atoms, 
$\mathcal{N}'$ can be $\Delta'^{++}$ particle with $Z'=2$.  

The effect of the $e'$ screening will be negligible if the mirror atom is not
too compact, i.e. the inverse radius of the mirror atom $1/a' \simeq \alpha m'_e$ 
is smaller than the transfer momentum $q= \sqrt{2M_A E_R}$, 
where $M_A$ is the mass of target atom and $E_R$ is recoil energy. 
In particular, for Na target in DAMA, 
considering that the relevant recoil energy range  is 2--6 keV electron equivalent (keVee) 
which corresponds to $E_R \simeq 6-20$ keV 
when one takes into account the quenching factor, we have $q>20$ MeV.  
Thus, for $m'_e < 1$ GeV the condition $1/q < a'$ is safely satisfied,  
and as  a consequence, the cross section 
depends on the recoil energy, $E_R$, and 
the relative velocity, $v$, as $d\sigma/dE_{R}\sim 1/(E_{R}v)^{2}$. 

Certainly, one could consider also contact interactions between ordinary and mirror nucleons 
due to interaction terms like $\frac{1}{M} \bar q \gamma_\mu q \bar q' \gamma_\mu q'$ 
between ordinary and mirror quarks which can be mediated by extra gauge 
bosons connecting two sectors, as e.g. of flavor symmetry
\cite{Berezhiani:1996ii}. 
However, if these interactions have cross-sections large enough for the dark matter direct 
detection, then the same interactions would bring two sectors into equilibrium in the early 
universe, violating the BBN limit and even overclosing the universe due to large masses 
of the sterile mirror  neutrinos. This can be avoided only by assuming a very low 
reheating temperature after the inflation. As for the photon-mirror photon 
kinetic mixing portal, the cosmological constraints are easily satisfied without any 
exotic assumptions on the post-inflationary reheating of the universe. 
By the similar reasons, this interaction portal avoids the experimental 
limits on the dark matter production  at the LHC and other accelerators.

\subsection{Interaction Rate}

The low-energy differential cross-section of the interaction between mirror and ordinary nuclei has the Rutherford-like form:
\be{dsigmaR}
\frac{d\sigma_{A,A'}
}{dE_{R}}=\frac{\mathcal{C}_{A,A'}}{E_{R}^{2}v^{2}}
\ee
where $E_{R}$ is the energy of the ordinary nucleus recoil, $v=|\vect{v}|$ is the relative velocity between the mirror nucleus and the ordinary one,
and:
\be{C}
\mathcal{C}_{A,A'}=\frac{2\pi \epsilon^{2}\alpha^{2}Z^{2}Z'^{2}}{M_{A}}\mathcal{F}^2_{A}\mathcal{F}^2_{A'}
\ee
where $\alpha$ is the fine structure constant, $Z$ and $Z'$ are the charge numbers of the ordinary and mirror nuclei,
$M_{A}$ is the mass of the ordinary nucleus, and  
$\mathcal{F}_{X}(qr_{X})$ $(X=A,A')$ are the Form-factors of ordinary and mirror nuclei, which depend on the momentum transfer, $q$,
and on the radius of $X$ nucleus.

As a consequence the differential interaction rate of mirror nuclei on a target composed by more than one kind of nucleus is: 
\be{RateR}
\frac{d\mathcal{R}}{dE_{R}}=\sum_{A,A'}N_{A}n_{A'}\int 
\frac{d\sigma_{A,A'}}{dE_{R}} F_{A'}(\vect{v},\vect{v}_{E}) v d^{3}v = \ee
$$= \sum_{A,A'}N_{A}n_{A'}\frac{\mathcal{C}_{A,A'}}{E_{R}^{2}}\int_{v>v_{min}(E_{R})}
\frac{F_{A'}(\vect{v},\vect{v}_{E})}{v}d^3v,$$
where: i) $N_{A}$ is the number of the target atoms of specie $A$ per kg of detector; ii) 
$n_{A'}=\rho_{dm}f_{A'}/M_{A'}$ with $\rho_{dm}$ local dark matter density, $f_{A'}$ fraction of the
specie $A'$ in the dark halo, and $M_{A'}$ mirror nucleus mass; iii) the sum is performed over the mirror nuclei involved in the interactions
($A'$) and over the target nuclei in the detector ($A$); iv) $F_{A'}(\vect{v},\vect{v}_{E})$ is the velocity distribution of the $A'$ mirror 
nuclei in the laboratory frame, which depends on the velocity of the Earth in the galactic frame:
$\vect{v}_{E}$.
The lower velocity limit 
$v_{min}(E_R)$ is
\be{vmin}
v_{min}(E_R)=\sqrt{\frac{(M_{A}+M_{A'})^{2}E_{R}}{2M_{A}M_{A'}^2}}.
\ee
As mentioned above, in our model we consider just one specie  of mirror nuclei. 
Our benchmark model is the mirror hydrogen ($A'=Z'=1$), with mass  $M_{A'}\simeq 5$~GeV 
and fraction $f_{A'} = f$. 
Alternatively, one can consider the helium like $\Delta$-atom that we have discussed above, 
with $A'=1, Z'=2$ and with mass again $M_{A'}\simeq 5$~GeV. All numerical 
results, presented below in the case of mirror hydrogen in terms of $\epsilon f^{1/2}$, 
would be equivalent  to $Z' \epsilon f^{1/2}$ in the case of mirror nuclei with $Z'>1$  with 
the same mass. So, for $\Delta$-atom one just puts  $Z'=2$.  

In order to compare the theoretical differential rate with the experimental data,
one has to take into account the detector response 
by means of the $\mathcal{K_A}(E|E_{R})$ kernel for each $A$ nucleus in the target-detector; $E$ is the detected energy
in keV electron equivalent (generally in literature indicated simply as keV).
Thus, the theoretical differential counting rate can be written as:
\be{RateG}  
\frac{d\mathcal{R}}{dE}=\sum_A \int \mathcal{K_A}(E|E_{R})\frac{d\mathcal{R_A}}{dE_{R}}dE_{R}.
\ee
where $\frac{d\mathcal{R_A}}{dE_{R}}$ is the differential interaction rate on the $A$ nucleus in the detector.
The $\mathcal{K_A}(E|E_{R})$ kernel accounts for the detector's energy resolution (generally through a gaussian convolution) and for the
transformation of the  nuclear recoil energy in keV electron equivalent through the use of a quenching factor. For a discussion about the 
quenching factors see later. 
In particular, that kernel can be written as:
\be{kernelK}
\mathcal{K_A}(E|E_{R})=\int \mathcal{G}(E|E') \mathcal{Q_A}(E'|E_{R}) dE'
\ee
where: 
$$\mathcal{G}(E|E') = \frac{1}{\sqrt{2\pi}\sigma}e^{-(E-E')^2/2\sigma^2}$$
takes into account the energy resolution $\sigma$ (generally function of $E'$) by a gaussian behaviour, and
$\mathcal{Q_A}(E'|E_{R})$ takes into account the energy transformation through the quenching factor (see later).  
For example, the latter kernel can be written in the simplest case of a constant quenching factor $q_A$ as:
$\mathcal{Q_A}(E'|E_{R})= \delta(E'-q_AE_R)$.
 
The expected differential rate depends 
on the Earth's velocity in the galactic frame, $\vect{v}_{E}$, which depends on the time of the year.
Projecting $\vect{v}_E(t)$ on the galactic plane, one can write:
$v_E(t) = v_{\odot} + v_{\oplus} cos\gamma cos\omega(t-t_0)$;
here $v_{\odot}$ is the Sun's velocity with respect
to the galactic halo ($v_{\odot} \simeq v_0 + 12$ km/s and $v_0$ is
the local velocity),
$v_{\oplus} \simeq$ 30 km/s is the Earth's orbital
velocity around the Sun on a plane with inclination
$\gamma$ = 60$^o$ with respect to the galactic plane. Furthermore,
$\omega$= 2$\pi$/T with T=1 year and roughly $t_0$ $\simeq$ June 2$^{nd}$
(when the Earth's speed in the galactic halo is at maximum). The Earth's velocity can be
conveniently expressed in unit of
 $v_0$: $\eta(t) = v_E(t)/v_0 = \eta_0 +
 \Delta\eta cos\omega(t-t_0)$,
where -- depending on the assumed value of the
local velocity ranging between 170 and 270 km/s -- $\eta_0$=1.04-1.07 is the yearly average of $\eta$ and
$\Delta\eta$ = 0.05-0.09. Since $\Delta\eta\ll\eta_0$, the expected counting
rate can be expressed by the first order Taylor expansion:
\begin{equation}
\frac{d\mathcal{R}}{dE}[\eta(t)] = \frac{d\mathcal{R}}{dE}[\eta_0] +
\frac{\partial}{\partial \eta} \left( \frac{d\mathcal{R}}{dE} \right)_{\eta =
\eta_0} \Delta \eta \cos\omega(t - t_0) .
\end{equation}
Averaging this expression in a given energy interval one obtains:
\begin{equation}
        \mathcal{S}\lbrack\eta(t)\rbrack = \mathcal{S}\lbrack\eta_0\rbrack
  + \left[\frac{\partial  \mathcal{S}}{\partial \eta}\right]_{\eta_0}
\Delta\eta cos\omega(t-t_0) =\mathcal{S}_0 + \mathcal{S}_m cos\omega(t-t_0),
\label{eq:sm}
\end{equation}
with the contribution from the highest order terms less than 0.1$\%$;
$\mathcal{S}_{m}$ and $\mathcal{S}_{0}$ are the modulated and the unmodulated
part of the expected differential counting rate, respectively.

\section{Details of the analysis}

The data analysis in the Mirror DM model framework considered here allows the determination of the
$\sqrt{f}\epsilon$ parameter. As mentioned this corollary analysis is model dependent; thus, it is 
important to point out at least the main topics which enter in the $\sqrt{f}\epsilon$ determination 
and the related uncertainties. In the following the main ones are addressed.

\subsection{Phase-space distribution functions of the dark halo}
\label{halo}

In order to derive the $\sqrt{f}\epsilon$ parameter 
a specific phase-space Distribution Function (DF) of the Mirror Dark Matter
in the galactic halo has to be adopted. A large number of possibilities is available in literature; 
thus, this introduces large uncertainties in the predicted 
theoretical rate. In addition, 
it is also possible the presence of non-virialized components, as
 streams in the dark halo coming from external sources with respect to our galaxy \cite{32}; these latter possibilities 
are not included in the present analysis.
 In conclusion,  it is strongly limiting/arbitrary to just consider 
 an isothermal profile\footnote{It is also worth noting 
that the isothermal halo is an unphysical model; for example, the mass would diverge and one has to 
adopt a by-hand cut-off. Let us remark, however, that flat density profile for the Galaxy within 
the radius of 10 kpc can be obtained if the DM particles have self-interaction cross-section   
$\sigma/M \sim 10^{-24}-10^{-23}~{\rm cm}^2/{\rm GeV}$
\cite{Spergel:1999mh,Wandelt:2000ad}. } 
with local 
 parameters $v_{0}=220\, \rm km/s$ and $\rho_{dm}\simeq 0.3\, \rm GeV/cm^{3}$
 without taking in consideration other existing possibilities in the distribution of 
 velocity and spatial coordinates permitted by astrophysical observations.

In this paper, we will consider a large (but not exhaustive) class of dark halo models,  
as already done in previous analyses for other DM candidates \cite{allDM8,RNC,ijmd,ijma,epj06,ijma07,chan,wimpele,ldm,bot11,review}; 
these models are summarized in Table \ref{tb:halo}.
An extensive discussion about
some of the more credited realistic halo models has been reported in Ref. \cite{allDM8,RNC}.
In particular, the considered classes of halo models correspond to: i) spherically symmetric matter density with
isotropic velocity dispersion (A); ii) spherically symmetric matter density with
non-isotropic velocity dispersion (B); iii) axisymmetric models (C);
iv) triaxial models (D);
v) moreover, in the case of axisymmetric models it is possible to include either an halo co-rotation
or an halo counter-rotation.
\begin{table}[!hbt]
\begin{center}
\vspace{-0.3cm}
\caption{
  Summary of the considered consistent halo models \cite{allDM8,RNC}.
  The labels in the first column identify the models.
  In the third column the values of the related considered parameters are reported \cite{allDM8,RNC};
  other choices are also possible as well as other halo models.
  The models of the Class C have also been considered including
  possible co--rotation and counter-rotation of the dark halo.
}
\begin{tabular}{|c|l|c|}
\hline\hline
\multicolumn{3}{|l|}{{\bf Class A:  spherical $\bf \rho_{dm}$,
isotropic velocity dispersion}} \\   
\hline
A0 & {\rm ~Isothermal Sphere}   &     \\
A1 & {\rm ~Evans' logarithmic}  & $R_c=5$ kpc \\
A2 & {\rm ~Evans' power-law}  & $R_c=16$ kpc, $\beta=0.7$ \\
A3 & {\rm ~Evans' power-law}  & $R_c=2$ kpc, $\beta=-0.1$ \\
A4 & {\rm ~Jaffe}               & $\alpha=1$, $\beta=4$,
$\gamma=2$, $a=160$ kpc \\
A5 & {\rm ~NFW}                  & $\alpha=1$, $\beta=3$,
$\gamma=1$, $a=20$ kpc \\
A6 & {\rm ~Moore et al.}    & $\alpha=1.5$, $\beta=3$,
$\gamma=1.5$, $a=28$ kpc  \\
A7 & {\rm ~Kravtsov et al.} & $\alpha=2$, $\beta=3$,
$\gamma=0.4$, $a=10$ kpc   \\
\hline
\multicolumn{3}{|l|}{{\bf Class B: spherical $\bf \rho_{dm}$,
non--isotropic velocity dispersion    }} \\
\multicolumn{3}{|l|}{{\bf (Osipkov--Merrit, $\bf \beta_0=0.4$)}} \\
\hline
B1 & {\rm ~Evans' logarithmic} & $R_c=5$ kpc \\
B2 & {\rm ~Evans' power-law} & $R_c=16$ kpc, $\beta=0.7$  \\
B3 & {\rm ~Evans' power-law} & $R_c=2$ kpc, $\beta=-0.1$  \\
B4 & {\rm ~Jaffe}           & $\alpha=1$, $\beta=4$,
$\gamma=2$, $a=160$ kpc  \\
B5 & {\rm ~NFW}             & $\alpha=1$, $\beta=3$,
$\gamma=1$, $a=20$ kpc   \\
B6 & {\rm ~Moore et al.}    & $\alpha=1.5$, $\beta=3$,
$\gamma=1.5$, $a=28$ kpc   \\
B7 & {\rm ~Kravtsov et al.} &  $\alpha=2$, $\beta=3$,
$\gamma=0.4$, $a=10$ kpc    \\
\hline
\multicolumn{3}{|l|}{{\bf Class C:  Axisymmetric $\bf \rho_{dm}$}} \\
\hline
C1 & {\rm ~Evans' logarithmic} & $R_c=0$, $q=1/\sqrt{2}$ \\
C2 & {\rm ~Evans' logarithmic} & $R_c=5$ kpc, $q=1/\sqrt{2}$ \\
C3 & {\rm ~Evans' power-law} & $R_c=16$ kpc, $q=0.95$, $\beta=0.9$ \\
C4 & {\rm ~Evans' power-law} & $R_c=2$ kpc, $q=1/\sqrt{2}$, $\beta=-0.1$\\
\hline
\multicolumn{3}{|l|}{{\bf Class D: Triaxial $\bf \rho_{dm}$
  ($\bf q=0.8$, $\bf p=0.9$)}} \\
\hline
D1 & {\rm ~Earth on maj. axis, rad. anis.}    & $\delta=-1.78$  \\
D2 & {\rm ~Earth on maj. axis, tang. anis. }    &   $\delta=16$ \\
D3 & {\rm ~Earth on interm. axis, rad. anis.}  &  $\delta=-1.78$ \\
D4 & {\rm ~Earth on interm. axis, tang. anis.} & $\delta=16$ \\
\hline\hline
\end{tabular}
\label{tb:halo}
\end{center}
\end{table}

In our analysis we also consider the physical ranges of the main halo parameters: the local total DM density
$\rho_{dm}$ and the local velocity $v_{0}$ as previously discussed in Ref. \cite{allDM8}.
In particular the range of the possible $v_{0}$ value is from $170\, \rm km/s$ to $270\, \rm km/s$.
We will consider for $\rho_{dm}$, its minimal value $\rho_{dm}^{min}$ or its maximal one $\rho_{dm}^{max}$,
compatible with the given $v_{0}$. The $\rho_{dm}^{min}$ and $\rho_{dm}^{max}$ are defined (as in Ref. \cite{allDM8})
as the values, associated to a specific $\rho_{dm}$, which provide a visible mass maximal and minimal contribution, respectively, to the 
total mass of the dark halo compatible with the astrophysical observations and constraints.
The particular values for $\rho_{dm}^{min},\rho_{dm}^{max}$ are related to the particular DF and the particular
$v_{0}$ considered. See Table III of Ref. \cite{allDM8} for
maximal densities at given $v_{0}$ and given model; for $v_{0}$= 170 km/s $\rho_{dm}$ ranges from 0.17 to 0.67 GeV cm$^{-3}$, while for 
$v_{0}$= 220 km/s $\rho_{dm}$ ranges from 0.29 to 1.11 GeV cm$^{-3}$, and for $v_{0}$= 270 km/s $\rho_{dm}$ ranges from 0.45 to 1.68 GeV cm$^{-3}$, 
depending on the halo model.
Finally, we consider a DM escape velocity from the galactic gravitational potential $v_{esc}=650\, \rm km/s$,
as often considered in literature; however, it is also affected by significant uncertainty. No sizeable differences
are observed in the final results when a different value of $v_{esc}=550\, \rm km/s$ is considered.

\subsection{Nuclei and Dark Matter Form factors}

Other important items for the determination of the expected signal counting rate are the nuclei and DM form factors.
Usually a {\it Helm form factor} \cite{Helm1,Helm2} is considered\footnote{It should be noted that 
the Helm form factor is the less favorable one e.g. for iodine and requires larger SI cross-sections for 
a given signal rate; in case 
other form factor profiles, considered in the literature, would be used \cite{RNC}, the allowed 
parameters space would extend.} for each X ordinary and mirror nucleus: 
\be{Helm}
\mathcal{F}_{X}(q r_{X})=3\frac{j_{1}(q r_{X})}{q r_{X}}e^{-(qs)^{2}/2}
\ee
where $q=(2M_{X}E_{R})^{1/2}$ is the momentum transfer, $r_{X}$ is the effective nuclear radius
(the normalization in natural units $\hbar=c=1$ is understood), $s$ is the nuclear surface thickness, and
$j_{1}$ is the Bessel function of order 1. 
This analytical expression is sufficiently good for our purposes, especially comparing the uncertainties coming e.g. from the astrophysical side.
We consider here $ s\simeq 1$ fm, $r_{X}=\sqrt{r^2_0-5s^2}$ and $r_0=1.2 A^{1/3}$ fm;
in case of light nuclei, as e.g. the mirror Hydrogen is, we use $ s\simeq 0.9$ fm and $r_{X}=1.14 A^{1/3}$ fm. 
However, let us note that the mirror proton is even more compact that ordinary proton 
and thus we can safely take the nuclear form factor equal to one.
In the analysis some uncertainties on the nuclear radius and on the nuclear surface thickness parameters in the Helm SI form factors 
have been included (see e.g. \cite{RNC,bot11}).

\subsection{Quenching factors and Channeling effect}

In the present analysis, we consider with the appropriate care the uncertainties
in the quenching factors. A precise experimental determinations of these quantities
are difficult for all kind of detectors. 
In fact, generally the direct measurements of quenching factors are 
performed with reference detectors, and -- in some cases -- 
with reference detectors having features quite different from the ones used in the underground running conditions;
in other cases the
quenching factors are not even measured at all. In addition it should be noted that the quenching factor value is a feature of each 
specific detector and not a fixed property of a given material. Moreover, the real nature of these measurements and the used neutron beam/sources 
may not point out all the possible contributions or instead may cause uncertainties because e.g. of the presence of spurious 
effects due to interactions with dead materials as e.g. housing or
cryogenic assembling, if any; therefore, they are intrinsically more uncertain than generally derived. 
A discussion dedicated to the case of Na and I quenching factors in DAMA experiments has been given in section II of Ref. \cite{bot11};
analogous / similar discussions should be pursued for every other case. In fact,
the related uncertainties affect all the results both in terms of exclusion 
plots and in terms of allowed regions/volumes; thus, comparisons with a fixed set of assumptions and parameters values are intrinsically strongly 
uncertain. 

According to Ref. \cite{bot11,ldm}, in the present analysis three possibilities for the Na and I quenching factors 
have been considered: Q$_I$) the quenching factors of Na and I ``constants''
with respect to the recoil energy $E_{R}$: $q_{Na}\simeq 0.3$ and $q_{I}\simeq 0.09$ as measured by 
DAMA with neutron source integrated over the $6.5-97\, \rm keV$ and the $22-330\, \rm keV$ 
recoil energy range, respectively \cite{allDM1}; Q$_{II}$) the quenching factors evaluated as in Ref. \cite{Tretyak}
varying as a function of $E_R$; Q$_{III}$) the quenching factors with the same behaviour of Ref. \cite{Tretyak},
but normalized in order to have their mean values  
consistent with Q$_{I}$ in the energy range considered there.
 
Another important effect is the {\it channeling} of low energy ions along axes and planes of the NaI(Tl) DAMA crystals.
This effect can lead to an important deviation, in addition to the other uncertainties discussed above. In fact, the 
{\it channeling} effect in crystals implies that a fraction of nuclear recoils are channeled and experience
much larger quenching factors than those derived from neutron calibration (see \cite{bot11,chan} for a discussion
of these aspects). 
Since the {\it channeling} effect cannot be generally pointed out with neutron measurements as
already discussed in details in Ref. \cite{chan}, only modeling has been produced up to now. In particular,
the modeling of the {\it channeling} effect described by DAMA in Ref. \cite{chan} is able to reproduce the recoil spectrum
measured at neutron beam by some other groups (see Ref. \cite{chan} for details). 
For completeness, we mention an alternative {\it channeling} model, as that of Ref. \cite{Mat08}, where larger probabilities of the
planar channeling are expected. Moreover, we mention the analytic calculation claiming that the {\it channeling} effect holds
for recoils coming from outside a crystal and not from recoils produced inside it, due to the blocking effect \cite{gelmini}.
Nevertheless, although some amount of blocking effect could be present, the precise description of the 
crystal lattice with dopant and trace contaminants is quite difficult and analytical calculations require 
some simplifications which can affect the result.
Because of the difficulties of experimental measurements and of theoretical estimate of this {\it channeling} effect,
in the following it  will be either included or not in order to give idea on the related uncertainty.

\subsection{Migdal effect}

In case of low mass DM particles giving rise to nuclear recoils it is also necessary to account for the Migdal effect;
this effect is known since long time and is described both in dedicated papers \cite{1,2} and in
textbooks \cite{3,4}. It has also been recently addressed for the DM field in Ref. \cite{5,6,7}. This effect consists in the
ionization and the excitation of bound atomic electrons induced by the presence of a recoiling atomic nucleus.
A detailed discussion of its impact in the corollary analyses in terms of some DM candidates is given in Ref. \cite{ijma07}.
In this paper, the case of mirror nuclei interacting with the target nuclei is considered; thus,
since the recoiling nucleus can ``shake off'' some of the atomic electrons, an electromagnetic contribution is present together with a recoil 
signal. Since this contribution is not quenched, one can expect that this part (usually unaccounted) can play a role,
mainly when low mass DM candidates are considered; however, in the present case of mirror matter, one can expect second order corrections 
(of an order not exceeding $10^{-1}$ on the expected counting rate) when the Migdal effect is accounted for.

\subsection{Further uncertainties on parameters}
\label{setabc}

In the analysis here reported, some discrete cases 
are considered to account for the uncertainties on the measured quenching factors and on the 
parameters used in the nuclear form factors, as already done in previous analyses for other DM candidates. 
The first case (set A) is obtained
considering the mean values of the parameters of the used nuclear form factors \cite{RNC}
and of the quenching factors. The set B adopts the same procedure as in Refs. \cite{allDM6,allDM7},
by varying (i) the mean values of the measured $^{23}$Na and $^{127}$I quenching factors
up to +2 times the errors; (ii) the nuclear radius, $r_A$, and the nuclear surface thickness parameter, $s$,
in the form factor from their central values down to -20\%.
In the last case (set C) the Iodine nucleus parameters are fixed at the values of case B,
while for the Sodium nucleus one considers: (i) $^{23}$Na quenching factor at the lowest value measured
in literature; (ii) the nuclear radius, $r_A$, and the nuclear surface thickness parameter, $s$,
in the SI form factor from their central values up to +20\%.

\subsection{Analysis procedures}

As mentioned, the approach exploited by the DAMA experiments is a model-independent signature
with very peculiar features: the DM annual modulation signature.
In this case the experimental observable is not -- as in other experiments -- the constant part of the signal, $\mathcal{S}_0$, but its
modulation amplitude, $\mathcal{S}_m$, as a function of energy. This approach has several advantages; in particular,
the only background of interest is the one able to mimic the
signature, i.e. able to account for the whole observed modulation amplitude and to
simultaneously satisfy all
the many specific peculiarities of this signature. No background of this sort has been
found or suggested by anyone over more than a decade.
Thus, this approach does not require any identification of $\mathcal{S}_0$ from the total counting rate
in order to establish the presence of DM particles in the galactic halo.
Therefore, the DM annual modulation signature allows
one to overcome the large uncertainties associated to: i) the 
many data selections/subtractions/statistical-discrimination
procedures; ii) the modeling of surviving
background in keV region; iii) the {\it a priori} assumption on the nature, interaction type. etc. of
the DM particle(s), which are necessary in the experiments where the experimental observable is $\mathcal{S}_0$. 

When the DM annual modulation signature is applied, $\mathcal{S}_0$ can be worked out 
-- for each considered framework -- by corollary model dependent analysis
through a maximum likelihood analysis which also takes into account the energy behaviour of each detector.
However, for simplicity the allowed regions in the parameters space (e.g. $\sqrt{f}\epsilon$ in the DM model
described in this paper)  
can also be calculated by comparing -- for each $k-th$ energy bin -- the measured DM 
annual modulation amplitude $S_{m,k}^{exp}$ 
with the expectation in the considered framework, $S_{m,k}^{th}$. In this procedure one should remind that 
the measured counting rate in the cumulative energy spectrum -- given by the sum of the constant background 
contribution $b_k$ and of the constant part of the signal $S_{0,k}$ --
is about 1 cpd/kg/keV in the lowest energy bins; in particular, as discussed
e.g. in Ref. \cite{taupnoz}, this constant background is estimated to be not lower 
than $\sim$ 0.75 cpd/kg/keV in the 2-4 keV energy region; thus, an upper limit 
on $S_0$ of $\sim$ 0.25 cpd/kg/keV ($S_{0,max}$) is derived \footnote{
It is worth noting that not to account for this experimental fact is one of the many reasons 
of incorrect allowed regions put forward by most authors for the particular scenario they adopt.}.

To compare the expectations with the experimental results, the 
modulation amplitudes as function of energy \cite{modlibra3} have been considered. 
The energy bin used here is 1 keV and 
the experimental modulation amplitude in the $k-th$ bin is 
$S_{m,k}^{exp} \pm \sigma_{k}$.
We compute the $\chi^{2}$ quantity:
\be{chi2}
\chi^{2}=\sum_{k}\frac{(\mathcal{S}_{m,k}^{exp}-\mathcal{S}_{m,k}^{th})^{2}}{\sigma_{k}^{2}}+\frac{(\mathcal{S}_{0,max}-\mathcal{S}_{0,2-4}^{th})^{2}}{\sigma_{2-4}^{2}}
\Theta(\mathcal{S}_{0,2-4}^{th}-\mathcal{S}_{0,max})
\ee
where the second term encodes the experimental bound about the unmodulated part 
of the signal; here $\sigma_{2-4} \simeq 10^{-3}$ cpd/kg/keV,
$\Theta$ is the Heaviside function, and $\mathcal{S}_{0,2-4}^{th}$ is the average 
expected signal counting rate in the $(2-4)$ keV energy interval.
The sum in eq. (\ref{chi2}) runs from the software energy threshold (2 keV) to 20 keV.  Given the 
sharp decreasing shape of the expected signal for the candidate considered here,
the results are strongly driven by the data points in the $(2-4)$ keV energy interval.

The $\chi^2$ of eq. (\ref{chi2}) in the Mirror DM model considered here is function of only one parameter: $\sqrt{f}\epsilon$;
thus, we can define:
$$\Delta \chi^{2}\{\sqrt{f}\epsilon\}=\chi^{2}\{\sqrt{f}\epsilon\}-\chi^{2}\{\sqrt{f}\epsilon=0\}.$$
The $\Delta \chi^{2}$ is a $\chi^{2}$ with one degree of freedom and is used to determine the allowed interval of the $\sqrt{f}\epsilon$ parameter 
at $5\sigma$ from the {\it null signal hypothesis}.

\section{Results}

The data have been analysed by taking into account the uncertainties discussed in the previous sections;
they have been accounted for by evaluating the results for the various sets of parameters as summarized in Table 
\ref{tb:mir}. 
\begin{table}[!hbt]
\caption{Results on the $\sqrt{f}\epsilon$ parameter in the considered scenarios obtained by analysing the DAMA data in a mirror DM framework
as discussed in the text. For each scenario the best fit value of the $\sqrt{f}\epsilon$ parameter
and the relative allowed interval (corresponding to model providing the deeper $\Delta \chi^2$) 
are reported as well as the cumulative allowed interval for $\sqrt{f}\epsilon$ obtained when considering all the above mentioned models.
The allowed intervals identify the $\sqrt{f}\epsilon$ values corresponding to C.L. 
larger than $5\sigma$ from the {\it null hypothesis}, that is $\sqrt{f}\epsilon=0$. See text. 
}
\begin{center}
\vspace{-0.5cm}
\resizebox{\textwidth}{!}{ 
\begin{tabular}{|c|c|c|c|c|r|}\hline
Scenario & Quenching        & Channeling & Migdal &  $\sqrt{f}\epsilon$ best      &  $\sqrt{f}\epsilon$ interval  \\
         & Factor           &            &        &                               &   $(\times 10^{-9})$          \\
\hline
     &                  &            &        &                                            &                        \\
$a$  & $Q_{I}$ \cite{allDM1}    & no         & no     &  $4.45 \times 10^{-9}$ ($9.2 \sigma$ C.L.) &        1.86--4.52      \\
     &                  &            &        &                                            &  (all) 1.73--114.      \\
$b$  & $Q_{I}$ \cite{allDM1}    & yes        & no     &  $2.89 \times 10^{-9}$ ($9.3 \sigma$ C.L.) &        1.16--2.93      \\
     &                  &            &        &                                            &  (all) 0.77--9.72      \\
$c$  & $Q_{I}$ \cite{allDM1}    & no         & yes    &  $4.40 \times 10^{-9}$ ($9.2 \sigma$ C.L.) &        1.85--4.47      \\
     &                  &            &        &                                            &  (all) 1.72--107.      \\
$d$  & $Q_{II}$ \cite{Tretyak}   & no         & no     &  $2.44 \times 10^{-9}$ ($9.5 \sigma$ C.L.) &        1.03--2.48      \\
     &                  &            &        &                                            &  (all) 0.94--12.3      \\
$e$  & $Q_{III}$ \cite{Tretyak}-normalized& no & no     &  $5.18 \times 10^{-9}$ ($9.0 \sigma$ C.L.) &        2.24--5.26      \\
     &                  &            &        &                                            &  (all) 1.89--60.1      \\
\hline
\end{tabular}
}
\label{tb:mir}
\end{center}
\end{table}
In particular, five scenarios have been considered depending on: 
i) the adopted quenching factors; ii) either inclusion or not of the channeling effect; 
iii) either inclusion or not of the Migdal effect. For each scenario the halo models 
(138 models as discussed in Sect. \ref{halo})
and the relative uncertainties (the three sets described in Sect. \ref{setabc}) have been considered.
In Table \ref{tb:mir} for each scenario the best fit $\sqrt{f}\epsilon$ parameter 
corresponding to the model providing the deeper $\Delta \chi^{2}$ is reported; 
in addition, the allowed intervals of the $\sqrt{f}\epsilon$ parameter for the deeper $\Delta \chi^{2}$ model and
for all the considered models are reported as well. 
These allowed intervals identify the $\sqrt{f}\epsilon$ values corresponding to C.L. 
larger than $5\sigma$ from the {\it null hypothesis}, that is $\sqrt{f}\epsilon=0$. 

\begin{figure}[!ht]
\begin{center}
\vspace{-1.2cm}
\includegraphics [width=0.98\textwidth]{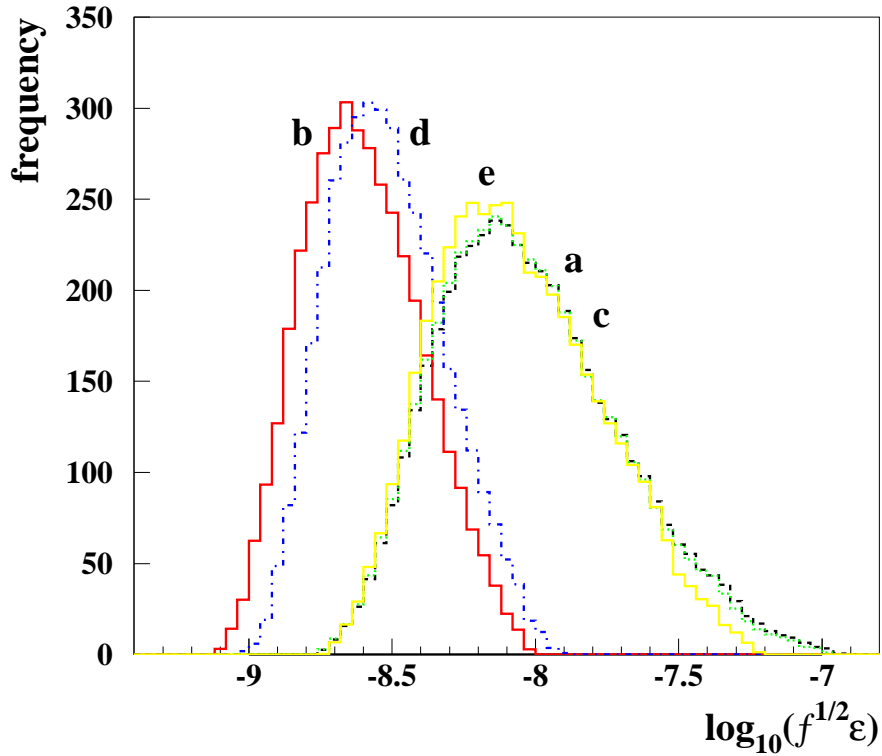}
\end{center}
\vspace{-0.8cm}
\caption{Distributions of the $\log_{10}(\sqrt{f}\epsilon)$ intervals obtained for all the models in each considered scenario,
identified by the letter, according to Table \ref{tb:mir}. See text.}
\label{fg:dist}   
\end{figure}

In Fig. \ref{fg:dist} the distributions of the $\log_{10}(\sqrt{f}\epsilon)$ allowed intervals 
of all the models are shown for each considered scenario. The scenarios $a$, $c$ and $e$ are very similar, while 
the scenarios either with channeling effect ($b$) or with with the $Q_{II}$ quenching factors ($d$)
support lower values of the Mirror DM coupling.

\begin{figure}[!pth]
\begin{center}
\includegraphics [width=11cm]{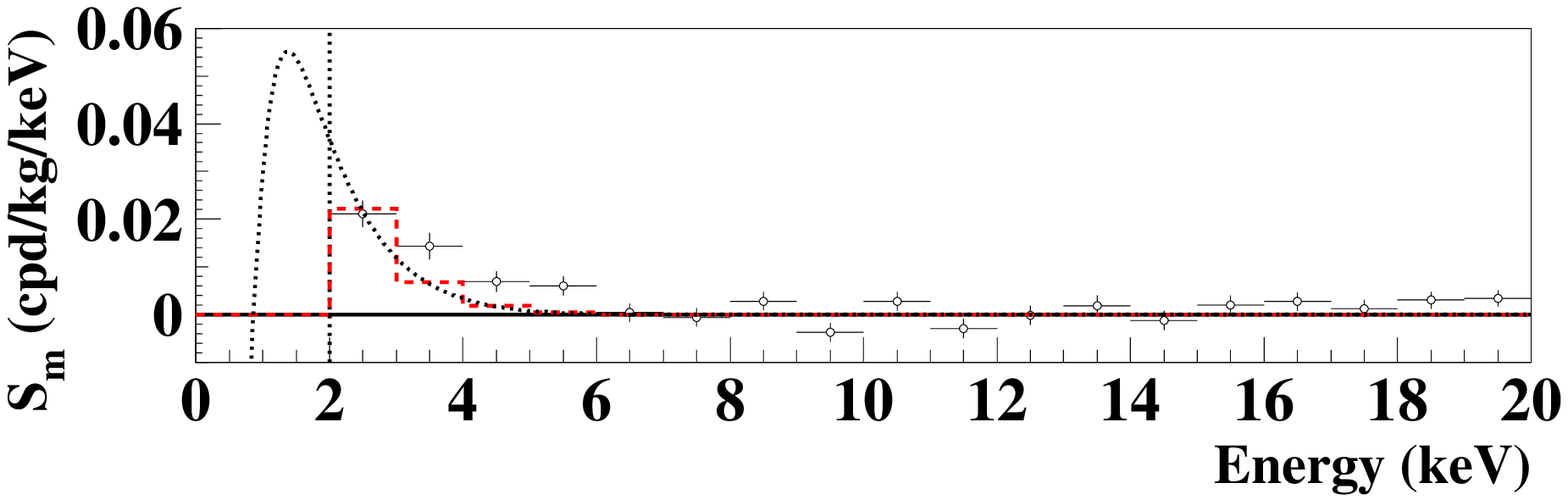} 
\includegraphics [width=11cm]{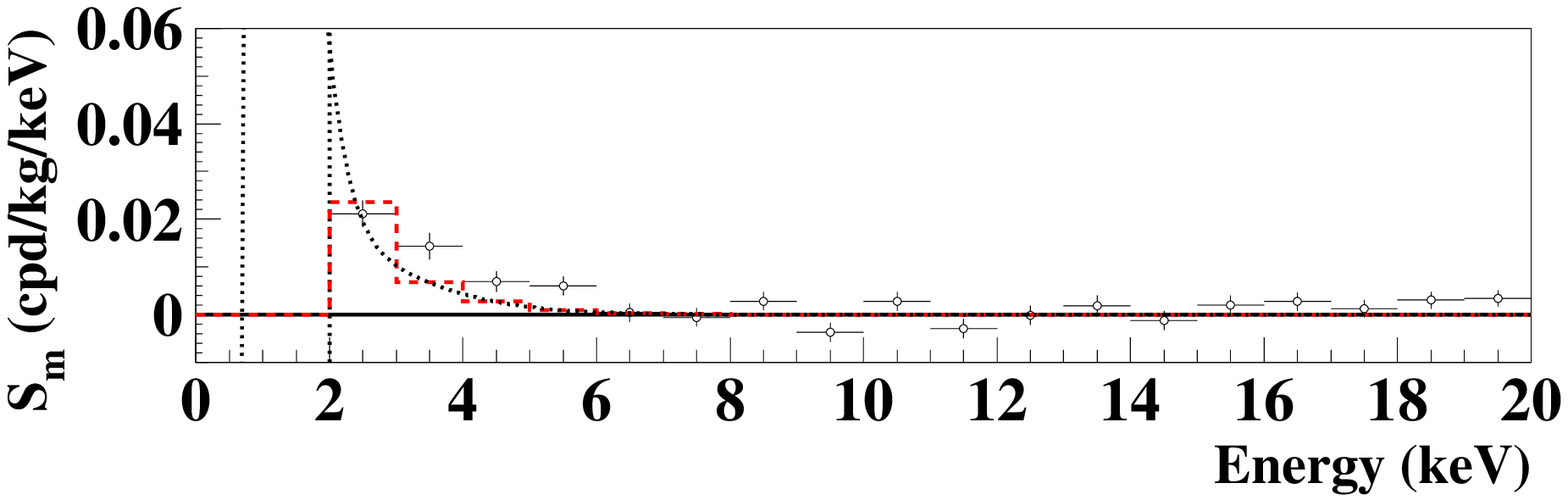} 
\includegraphics [width=11cm]{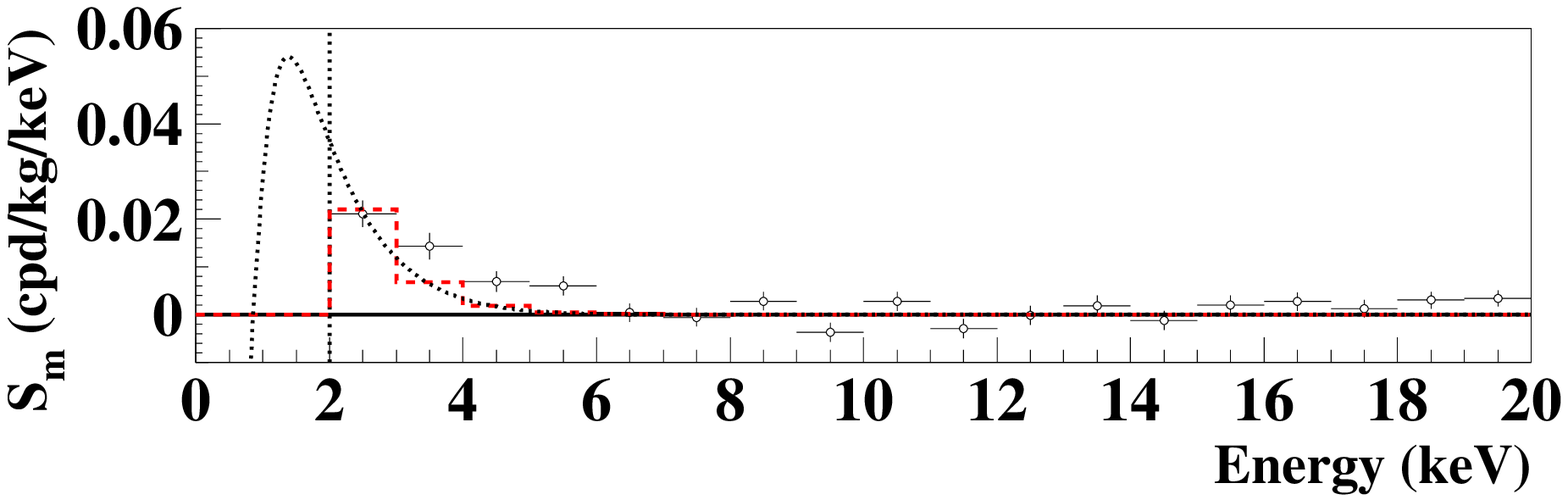} 
\includegraphics [width=11cm]{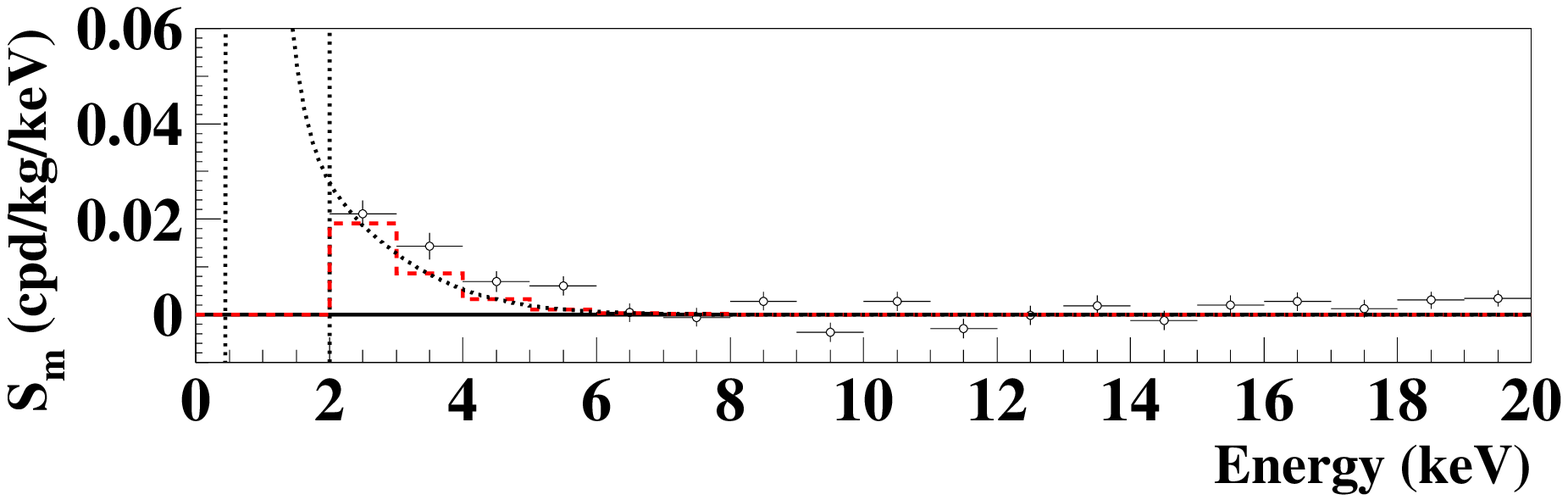} 
\includegraphics [width=11cm]{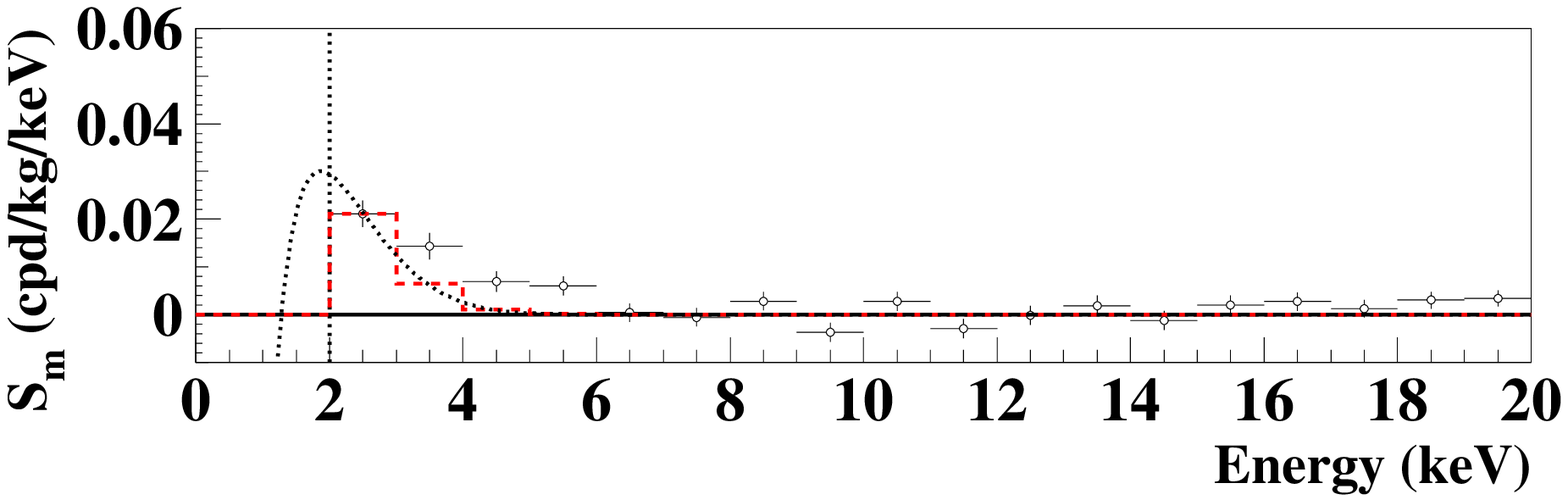} 
\end{center}
\vspace{-0.5cm}
\caption{Examples of expected modulation amplitude for the Mirror DM candidate considered here.
In particular, there are shown the best fit cases of the five scenarios (from top to bottom): 
i)    Mirror DM candidate with $Q_{I}$ quenching factors and without channeling effect (scenario $a$); 
ii)   Mirror DM candidate with $Q_{I}$ quenching factors and channeling effect included (scenario $b$); 
iii)  Mirror DM candidate with Migdal effect included in the interaction (scenario $c$);
iv)   Mirror DM candidate with $Q_{II}$ quenching factors (scenario $d$); 
v)    Mirror DM candidate with $Q_{III}$ quenching factors (scenario $e$).} 
\label{plot2}   
\end{figure}

In Fig. \ref{plot2} comparisons between the DAMA experimental modulation amplitudes and some expectations for
Mirror DM are shown.  
           
It is worth noting that in all the considered scenarios for mirror DM the DAMA signal in the 
2-6 keV energy interval arises from interactions mainly with Sodium nuclei \footnote{For example, 
the Iodine/Sodium contribution ratio in the best fit case of the scenario $d$ 
is 0.0062 and $2.4 \times 10^{-5}$
for the first two energy bins, (2--3) and (3--4) keV, respectively.}. This effect is due to the 
fact that the considered Mirror DM particle is quite light: $M_{A'}\simeq 5 m_{p}$.
            
\begin{figure}[!ht]
\centerline{\includegraphics [width=12cm]{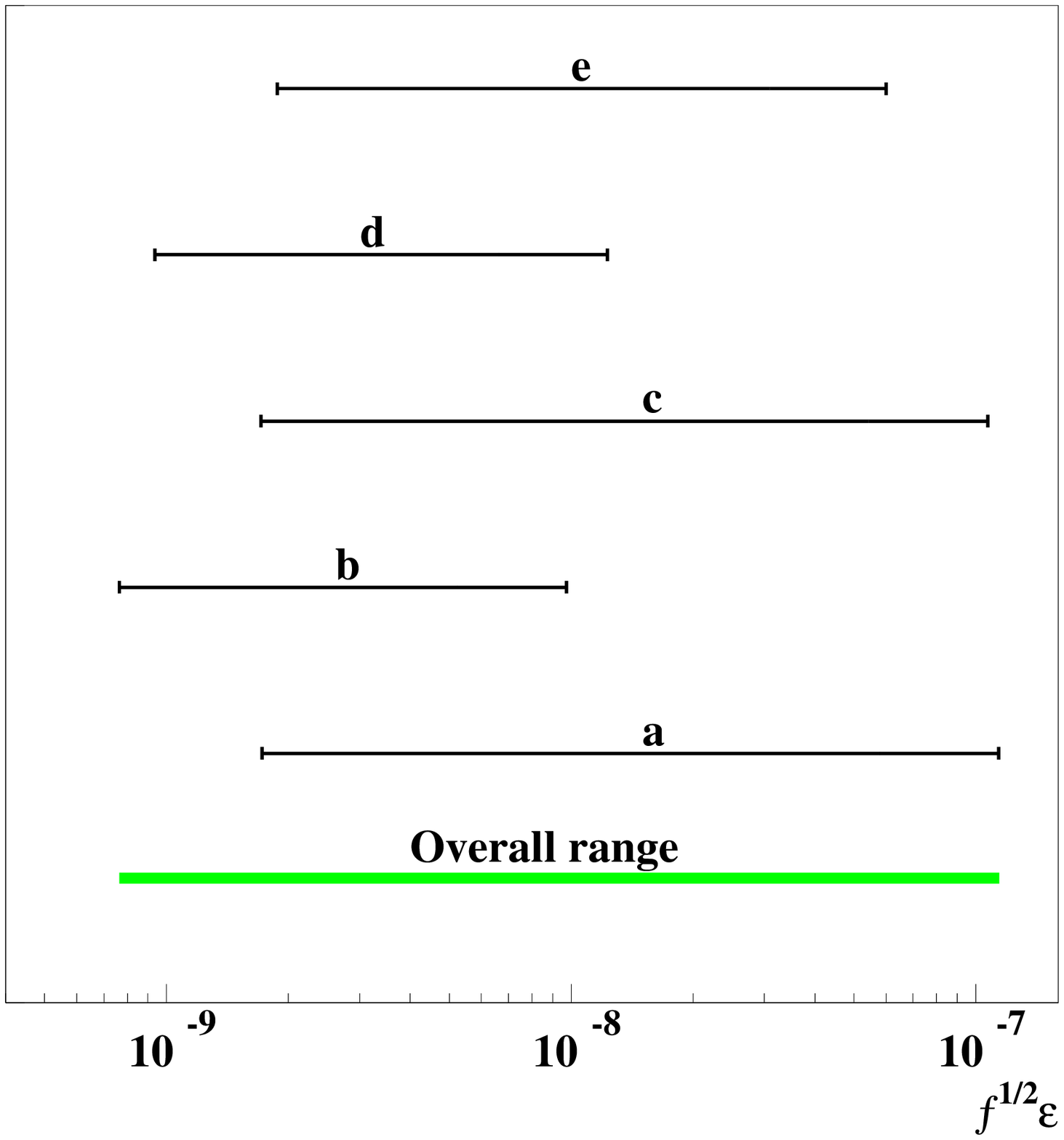} }
\caption{DAMA allowed intervals for the $\sqrt{f}\epsilon$ parameter, obtained by marginalizing all the models 
for each considered scenario as given in Table \ref{tb:mir}. The overall range is also reported.}
\label{plot3}   
\end{figure}

The cumulative allowed intervals of the $\sqrt{f}\epsilon$ parameter selected by the DAMA data for each scenario 
(see Table \ref{tb:mir}) are depicted in Fig. \ref{plot3},
where also the overall allowed band is shown.
The obtained values of the $\sqrt{f}\epsilon$ parameter
are well compatible with cosmological bounds cited in the introduction. 

Finally, we would like to note that the highest C.L. among all the analysed cases (see Table 
\ref{tb:mir}) is obtained for 
the scenario $d$ where: i) the quenching factors values are according to Ref. \cite{Tretyak};
ii) the halo model is the no-rotating Evans' logarithmic C2 model (see Table \ref{tb:halo}) with $v_0=270$ km/s, 
$\rho_{dm}=\rho^{max}_{dm}=1.68$ GeV cm$^{-3}$ and $v_{esc}=650\, \rm km/s$; 
iii) set B, as defined in Sect. \ref{setabc}. 

\begin{figure}[!ht]
\centerline{\includegraphics [width=12cm]{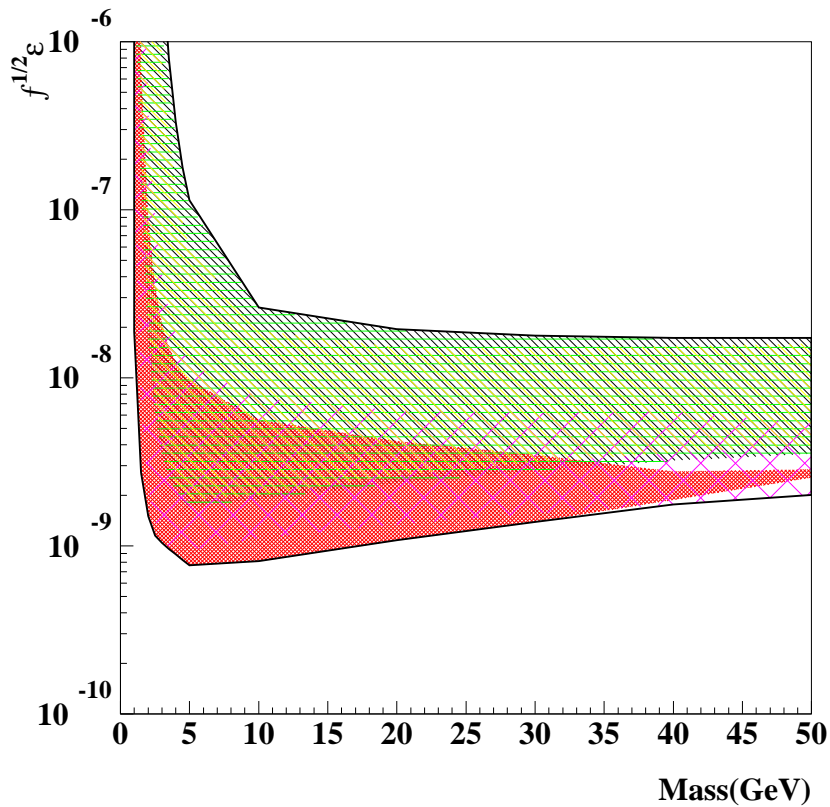} }
\caption{Allowed regions for the $\sqrt{f}\epsilon$ parameter 
as function of $M_{A'}$, when the assumption $M_{A'}\simeq 5 m_{p}$ is released,
obtained by marginalizing all the models for each considered scenario as given in Table \ref{tb:mir}.
The $M_{A'}$ interval from few GeV up to 50 GeV is explored. These allowed intervals identify the $\sqrt{f}\epsilon$ values corresponding to C.L. 
larger than $5\sigma$ from the {\it null hypothesis}, that is $\sqrt{f}\epsilon=0$.
The five scenarios defined in Table \ref{tb:mir} can be recognized on the basis of different hatching of the allowed regions;
the black line is the overall boundary.}
\label{fg:cont}   
\end{figure}

If the assumption $M_{A'}\simeq 5 m_{p}$ is released, the allowed regions for the $\sqrt{f}\epsilon$ parameter 
as function of $M_{A'}$ can be obtained by marginalizing all the models for each considered scenario as given in Table \ref{tb:mir}.
This is shown in Fig. \ref{fg:cont} where the $M_{A'}$ interval from few GeV up to 50 GeV is explored.
These allowed intervals identify the $\sqrt{f}\epsilon$ values corresponding to C.L. 
larger than $5\sigma$ from the {\it null hypothesis}, that is $\sqrt{f}\epsilon=0$.
The five scenarios defined in Table \ref{tb:mir} can be recognized on the basis of different hatching of the allowed regions;
the black line is the overall boundary.

\section{Conclusions}

The model-independent annual modulation effect measured by the DAMA Collaboration, 
which fulfills all the requirements of the DM annual modulation signature, 
has been examined in the context of asymmetric mirror matter model interacting 
with the ordinary nuclei via the photon-mirror photon kinetic mixing portal, 
$ \frac{\epsilon}{2} \, F^{\mu\nu} F'_{\mu\nu}$. We have assumed that mirror atoms 
constitute a fraction $f$ of the DM in the Galaxy, and we have derived 
the allowed physical intervals for the combination of parameters $\sqrt{f}\epsilon$, 
accounting also for various of the existing uncertainties. 

The allowed values for $\sqrt{f}\epsilon$ in the case of mirror hydrogen atom, $Z'=1$, 
ranges between $7.7 \times 10^{-10}$ to $1.1 \times 10^{-7}$. 
The values within this overall range are well compatible with cosmological bounds.
In particular, the best fit values among all the considered scenarios gives 
$\sqrt{f}\epsilon_{\rm b.f.} = 2.4 \times 10^{-9}$. 
If the assumption $M_{A'}\simeq 5$~GeV is relaxed, 
the allowed regions for the $\sqrt{f}\epsilon$ parameter 
as function of $M_{A'}$ have been obtained by marginalizing all the models for each considered scenario. 
We have also to remark that the atomic form-factor of mirror was not taken into 
account which is correct if mirror electron is light enough. In the case it is heavy, 
the obtained results for $\sqrt{f}\epsilon$ should be rescaled by the square root of the 
mirror atomic form-factor. In the case of helium-like $\Delta$-atoms with $Z'=2$, 
the ranges of $\sqrt{f}\epsilon$ obtained for the hydrogen case should be rescaled up 
by a factor 2. 

In our consideration, we did not take into account the possible modifications in the dark matter 
distribution in the Galaxy due to self-scattering properties of dark matter. 
However, marginalizing over a vast amount of models that we considered in our conservative 
approach should supersede all possible uncertainties that can be induced by this reason. 
Let us remark also that the photon-mirror photon kinetic mixing, besides the possibility 
of dark matter direct detection, could bring to other interesting cosmological consequences.
In particular, as it was shown in Ref. \cite{Berezhiani:2013dea}, 
in rotating protogalaxies the circular currents could emerge due the Rutherford-like scattering of the electrons with dark mirror particles in the dark matter halos,  
with cross-sections suppressed  by a parameter $\epsilon$.  
After a moderate dynamo amplification,  these currents could give origin 
to the observed galactic magnetic fields of few $\mu$G with the coherence scales of order of 1 kpc. 
Photon-mirror photon kinetic mixing portal and search for mini-charged particles in whole 
is elusive for LHC and other precision experiments are needed for detecting their 
production in laboratory conditions.  

\vspace{16mm} 
\noindent{\bf Acknowledgements} 
\vspace{2mm}

The work of A.A. and Z.B. was supported in part by the MIUR research 
grant ``Theoretical Astroparticle Physics'' PRIN 2012CPPYP7, and 
work of Z.B. was supported in part  by the 
Rustaveli National Science Foundation grant No. DI/8/6-100/12.

\end{document}